\documentclass[twocolumn,aps,prb,showpacs,preprintnumbers,amsmath,amssymb, superscriptaddress]{revtex4-1}

\usepackage{bm}
\usepackage{graphicx}
\usepackage{color,xspace}
\usepackage{ulem}

\renewcommand{\vec}{\mathbf}

\begin{document}

\title{Topologically entangled Rashba-split Shockley states on the surface of grey arsenic}

\author{Peng~Zhang}\thanks{These authors contributed equally to this work.}
\affiliation{Institute for Solid State Physics, University of Tokyo, Kashiwa, Chiba 277-8581, Japan}
\affiliation{Beijing National Laboratory for Condensed Matter Physics, and Institute of Physics, Chinese Academy of Sciences, Beijing 100190, China}
\author{J.-Z.~Ma}\thanks{These authors contributed equally to this work.}
\affiliation{Beijing National Laboratory for Condensed Matter Physics, and Institute of Physics, Chinese Academy of Sciences, Beijing 100190, China}
\author{Y.~Ishida}\thanks{These authors contributed equally to this work.}
\affiliation{Institute for Solid State Physics, University of Tokyo, Kashiwa, Chiba 277-8581, Japan}
\author{L.-X.~Zhao}\thanks{These authors contributed equally to this work.}
\affiliation{Beijing National Laboratory for Condensed Matter Physics, and Institute of Physics, Chinese Academy of Sciences, Beijing 100190, China}
\author{Q.-N.~Xu}
\affiliation{Beijing National Laboratory for Condensed Matter Physics, and Institute of Physics, Chinese Academy of Sciences, Beijing 100190, China}
\author{B.-Q.~Lv}
\affiliation{Beijing National Laboratory for Condensed Matter Physics, and Institute of Physics, Chinese Academy of Sciences, Beijing 100190, China}
\author{K.~Yaji}
\affiliation{Institute for Solid State Physics, University of Tokyo, Kashiwa, Chiba 277-8581, Japan}
\author{G.-F.~Chen}
\affiliation{Beijing National Laboratory for Condensed Matter Physics, and Institute of Physics, Chinese Academy of Sciences, Beijing 100190, China}
\affiliation{Collaborative Innovation Center of Quantum Matter, Beijing, China}
\author{H.-M.~Weng}
\affiliation{Beijing National Laboratory for Condensed Matter Physics, and Institute of Physics, Chinese Academy of Sciences, Beijing 100190, China}
\affiliation{Collaborative Innovation Center of Quantum Matter, Beijing, China}
\author{X.~Dai}
\affiliation{Beijing National Laboratory for Condensed Matter Physics, and Institute of Physics, Chinese Academy of Sciences, Beijing 100190, China}
\affiliation{Collaborative Innovation Center of Quantum Matter, Beijing, China}
\author{Z.~Fang}
\affiliation{Beijing National Laboratory for Condensed Matter Physics, and Institute of Physics, Chinese Academy of Sciences, Beijing 100190, China}
\affiliation{Collaborative Innovation Center of Quantum Matter, Beijing, China}
\author{X.-Q.~Chen}
\affiliation{Shenyang National Laboratory for Materials Science, Institute of Metal Research, Chinese Academy of Science, Shenyang 110016, China}
\author{L.~Fu}
\affiliation{Department of Physics, Massachusetts Institute of Technology, Cambridge, Massachusetts 02139, USA}
\author{T.~Qian}\email{tqian@iphy.ac.cn}
\affiliation{Beijing National Laboratory for Condensed Matter Physics, and Institute of Physics, Chinese Academy of Sciences, Beijing 100190, China}
\affiliation{Collaborative Innovation Center of Quantum Matter, Beijing, China}
\author{H.~Ding}\email{dingh@iphy.ac.cn}
\affiliation{Beijing National Laboratory for Condensed Matter Physics, and Institute of Physics, Chinese Academy of Sciences, Beijing 100190, China}
\affiliation{Collaborative Innovation Center of Quantum Matter, Beijing, China}
\author{S.~Shin}\email{shin@issp.u-tokyo.ac.jp}
\affiliation{Institute for Solid State Physics, University of Tokyo, Kashiwa, Chiba 277-8581, Japan}

\date{\today}

\begin{abstract}
\textbf{We discover a pair of spin-polarized surface bands on the (111) face of grey arsenic by using angle-resolved photoemission spectroscopy (ARPES). In the occupied side, the pair resembles typical nearly-free-electron Shockley states observed on noble-metal surfaces. However, pump-probe ARPES reveals that the spin-polarized pair traverses the bulk band gap and that the crossing of the pair at $\bar\Gamma$ is topologically unavoidable. First-principles calculations well reproduce the bands and their non-trivial topology; the calculations also support that the surface states are of Shockley type because they arise from a band inversion caused by crystal field. The results provide compelling evidence that topological Shockley states are realized on As(111). }
\end{abstract}


\maketitle

Due to the abrupt potential change on the crystal surface, the electronic states that only exist on the surface may appear. Shockely pointed out that such surface states, called Shockely states later, lie in an inverted bulk band gap induced by crystal field~\cite{ShockleyPR1939}. Shockley states were found on the surface of a variety of semiconductors and metals (Cu, Ag, Au, etc.)~\cite{KarlssonPRB1982, KevanPRB1987, JensenPRL1996, HufnerPRB2001, BaumbergerPRB2013}, generally exhibiting nearly-free-electron (NFE) band dispersion. Later, it was found that the surface states on Au(111) exhibit a spin-split doublet~\cite{JensenPRL1996}, due to the Rashba effect~\cite{Rashba1984}. As the topological surface states have attracted much interest recently, Yan \textit{et al}.\ predicted that the Shockley states on the surfaces of several noble metals are topologically entangled~\cite{FelserNC2015}. A direct measurement of the surface-to-bulk band connection by angle-resolved photoemission spectroscopy (ARPES) is necessary to confirm the band topology. However, it is difficult to determine their band topology by ARPES, because the surface bands of the noble metals connect to the bulk states at several electronvolts above the Fermi level ($E_\mathrm{F}$). 

In this work, we discover such topologically entangled Shockley states existing on the (111) face of grey arsenic by using time-resolved ARPES (TARPES) and spin-resolved ARPES (SARPES). In contrast with the other intensively studied group VA elemental crystals~\cite{PolliniJPCM1999,SatoPRB1999,HochstPRL2001, HofmannPRL2004, SasakiPRL2006, KanePRB2007, MatsudaARXIV2016, ChenArxiv2016}, such as bismuth, antimony, and black phosphorous, the electronic structures of arsenic crystals have not been investigated experimentally. We reveal that As(111) accommodates a pair of spin-split surface states. In the occupied states, the surface bands resemble the two-dimensional NFE bands subject to Rashba effect, the same as the Shockley states on the surfaces of noble metals. By visualizing the dispersions throughout the bulk band gap with TARPES, we find that the pair of the surface bands traverses the bulk band gap and is topologically entangled. Together with first-principles calculations, our results clearly show a nontrivial band topology of the Shockley states on As(111). 

High quality single crystals of grey arsenic were synthesized by the chemical vapor transport method. The TARPES measurements were carried out with 1.48-eV pump and 5.92-eV probe pulses at the repetition rate of 250~kHz~\cite{IshidaRSI2014}. The SARPES measurements were carried out using a custom-made ScientaOmicron DA30-L attached with twin VLEED spin detectors and a VUV laser system delivering 6.994-eV photons~\cite{ShinarXiv2016}. The synchrotron ARPES measurements were performed at the Dreamline of the Shanghai Synchrotron Radiation Facility and at the SIS beam line of the Swiss Light Source. For all the ARPES measurements, the samples were cleaved \textit{in situ} along the (111) plane and measured under high vacuum below 7 $\times$ 10$^{-11}$~Torr. The Vienna \textit{ab initio} simulation package (VASP) was employed for first-principles calculations. The generalized gradient approximation of Perdew-Burke-Ernzerhof type was used for the exchange-correlation potential. To calculate the As(111) surface states, we have built a lattice composed of 33 layers of As and a vacuum layer of 12~\AA. 

\begin{figure*}
\begin{center}
\includegraphics[width=0.9\textwidth]{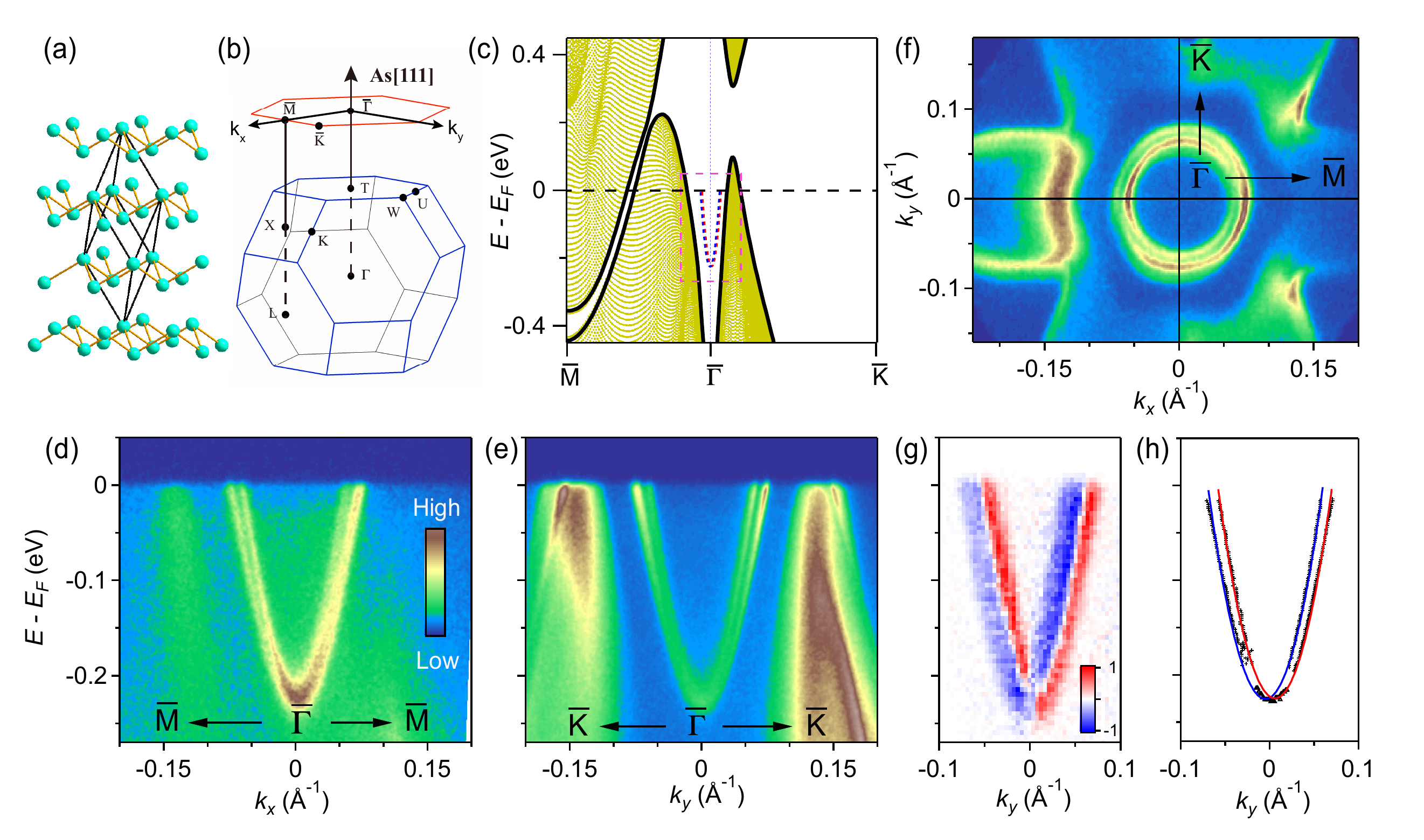}
\end{center}
 \caption{\label{band}  (a) Crystal structure of grey arsenic. Black lines indicate a rhombohedral primitive cell. (b) Bulk BZ and (111) surface BZ with indexes on the high-symmetry points. (c) Bulk bands projected on the (111) surface BZ. The pink dashed box indicates the energy and momentum ranges of panels (d) and (e). The two dashed curves are the fitted surface band dispersions in (h). (d) and (e) Band dispersions along $\bar\Gamma-\bar M$ and $\bar\Gamma-\bar K$, respectively, recorded with 7-eV laser.  (f) FS mapping recorded with 7-eV laser. (g) Spin-resolved energy-momentum intensity plot along $\bar\Gamma-\bar K$. The intensity scales the difference between spin-up and spin-down in the $x$ direction of the photoelectrons detected by the VLEED detector. (h) Band dispersions along $\bar\Gamma-\bar K$ fitted to a NFE model with Rashba splitting. The fittings use a combination of EDC (bottom part) and MDC peaks. }
\end{figure*}

Arsenic has three types of allotropes: grey, yellow, and black arsenic. In this work, we systematically investigate the electronic structure of grey arsenic, which has a rhombohedral primitive cell with space group of $R3m$ (No.~166). As illustrated in Fig.~\ref{band}(a), the crystal can be regarded as a stacking of As bilayer along the [111] direction. The As atoms within the bilayer form a buckled honeycomb lattice. The bulk Brillouin zone (BZ) and the (111) surface BZ are shown in Fig.~\ref{band}(b). The projection of the bulk bands on the (111) surface BZ, as shown in Fig.~\ref{band}(c), provides an easy view of the overall bulk band structure, which shows a semimetal feature, \textit{i.e.}, both the bulk valence and bulk conduction bands cross $E_\mathrm{F}$, while there is a gap between the two throughout the BZ.
The measured electronic states at $k \sim \pm $0.15 \AA$^{-1}$ in Figs.~\ref{band}(d) and \ref{band}(e) are consistent with the calculated bulk band structure. In addition, we observe a pair of parabolic bands, which splits along both $\bar\Gamma - \bar M$ and $\bar\Gamma - \bar K$ directions but degenerates at the $\bar\Gamma$ point [Figs.~\ref{band}(d) and \ref{band}(e)]. The parabolic bands form two concentric Fermi surfaces (FSs) enclosing the $\bar\Gamma$ point [Fig.~\ref{band}(f)]. By mapping the FS in the $k_y$-$k_z$ space with synchrotron radiations, we find that the Fermi momentum of the parabolic bands is located at $k_y \sim \pm$0.065~\AA$^{-1}$ throughout $k_z$ (White arrow in Fig.~S1 in the Supplemental Materials), confirming their surface nature.

The parabolic surface bands exhibit Rashba-type spin splitting, as revealed by the laser-SARPES measurements~\cite{ShinarXiv2016}. Figure~\ref{band}(g) displays that the bands along $\bar\Gamma-\bar K$ show significant spin polarization in the $x$ direction. In contrast, no obvious spin polarizations are observed in the $y$ and $z$ directions (see Fig.~S3 in the Supplemental Materials). Likewise, the bands along $\bar\Gamma-\bar M$ are mainly polarized along the $y$ direction (see Fig.~S2 in the Supplemental Materials). The spin texture is identical to that of the well-known Rashba-type spin-split surface states on Au(111)~\cite{JensenPRL1996}. Another consistency between the surface states on As(111) and Au(111) is that the surface bands are well described by the NFE model. In the NFE model, the band dispersions of Rashba-split states follow  $E(k) = \frac{\hbar^2}{2m^*}(k \pm \Delta k)^2 + E_0$, where $\Delta k$ describes the magnitude of the spin splitting and $m^*$ is the effective mass. Figure~\ref{band}(h) shows that the parabolic surface bands of As(111) are nicely described by the NFE model with $\Delta k \sim $ 0.0054~{\AA}$^{-1}$ and $m^* = 0.07~m_e$. 

\begin{figure*}
\begin{center}
\includegraphics[width=0.85\textwidth]{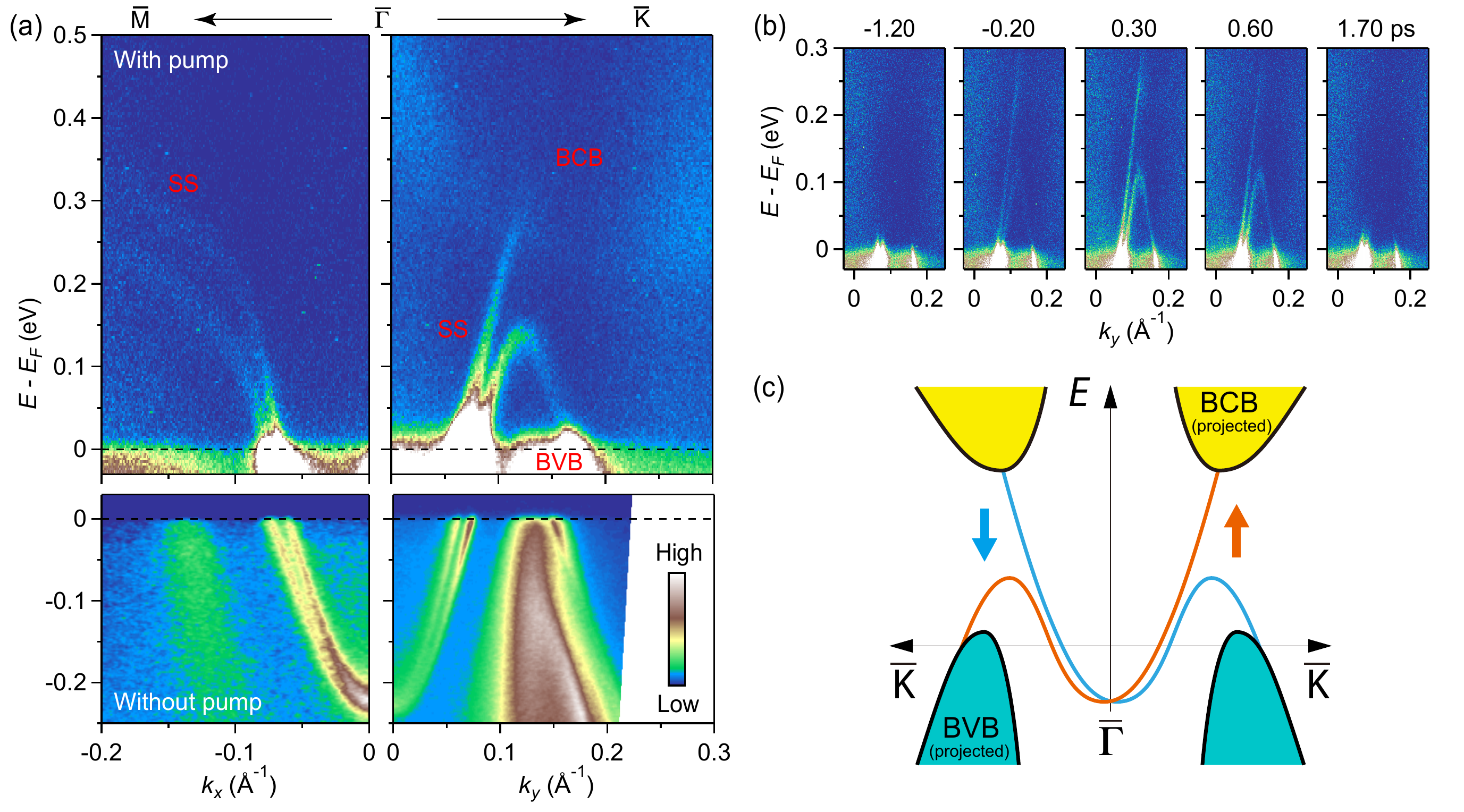}
\end{center}
 \caption{\label{tr} (a) Band dispersions below and above $E_\mathrm{F}$ along $\bar M-\bar\Gamma-\bar K$. Top panels: band dispersions above $E_\mathrm{F}$ recorded by TARPES measurements with a pump-probe method. Bottom panels: band dispersions below $E_\mathrm{F}$ recorded by conventional ARPES measurements with laser. (b) TARPES intensity snapshots along $\bar\Gamma-\bar K$ with various pump-probe delay times, showing the excitation/decay process of the states above $E_\mathrm{F}$. (c) Sketch of the topology of the surface and bulk band structures. Red and blue lines indicate the surface bands with opposite spin orientations. Yellow and green regions represent the projections of bulk conduction bands (BCB) and bulk valence bands (BVB), respectively. 
 }
\end{figure*}

While the occupied surface states on As(111) are well described by the NFE model, we clearly observe that the dispersions in the unoccupied side exhibit significant deviation along both $\bar\Gamma-\bar M$ and $\bar\Gamma-\bar K$ directions in Fig.~\ref{tr}(a).  The unoccupied states above $E_\mathrm{F}$ were revealed by TARPES with the pump-and-probe method: When impinged by an intense femtosecond pump pulse, electrons are redistributed into the unoccupied side, so that the bands therein can be observed by TARPES~\cite{IshidaRSI2014}. The pump-induced dynamics along the $\bar\Gamma - \bar K$ direction is displayed in Fig.~\ref{tr}(b). In the $\bar\Gamma - \bar K$ direction, one of the pair disperses into the conduction band, while the other turns back and merges into the valence band. As illustrated in Fig.~\ref{tr}(c), such a surface-to-bulk connection shows the nontrivial band topology. Along the $\bar\Gamma$-$\bar M$ direction, the topological connection between the surface and bulk states is not directly determined from the experimental data, but confirmed by first-principles calculations shown below. 

\begin{table}
\caption{Symmetry labels for the five valence bands of grey arsenic at eight TRIMs, which satisfy $\vec{k} = -\vec{k} + \vec{G}$ for a reciprocal lattice vector $\vec{G}$. The $Z_2$ topological index $\nu_0$ can be expressed as the parity product over all eight TRIMs, $(-1)^{\nu_0}= \prod_{i=1}^{8}\delta_i$, where $\delta_i$ is the parity product at each TRIM~\cite{KanePRB2007}.}
\centering
\begin{tabular}{lcccccc}
\hline\hline
TRIMs                                    &   Parities of occupied bands &   Product      \\
\hline
1 $\Gamma$                              &  $+$     $-$     $+$    $+$    $+$               &   $-$            \\
3 L              				&  $+$     $-$     $-$    $+$    $+$               &   $+$            \\
3 X   						&  $-$     $+$    $+$     $-$     $-$                &   $-$            \\
1 T                                    		&   $-$     $+$     $-$     $-$     $+$               &   $-$            \\
                                                   &                                           &   $Z_2(\nu_0 = 1)$ \\
\hline        
\end{tabular}
\label{table:parity}
\end{table}

\begin{figure}
\begin{center}
\includegraphics[width=0.5\textwidth]{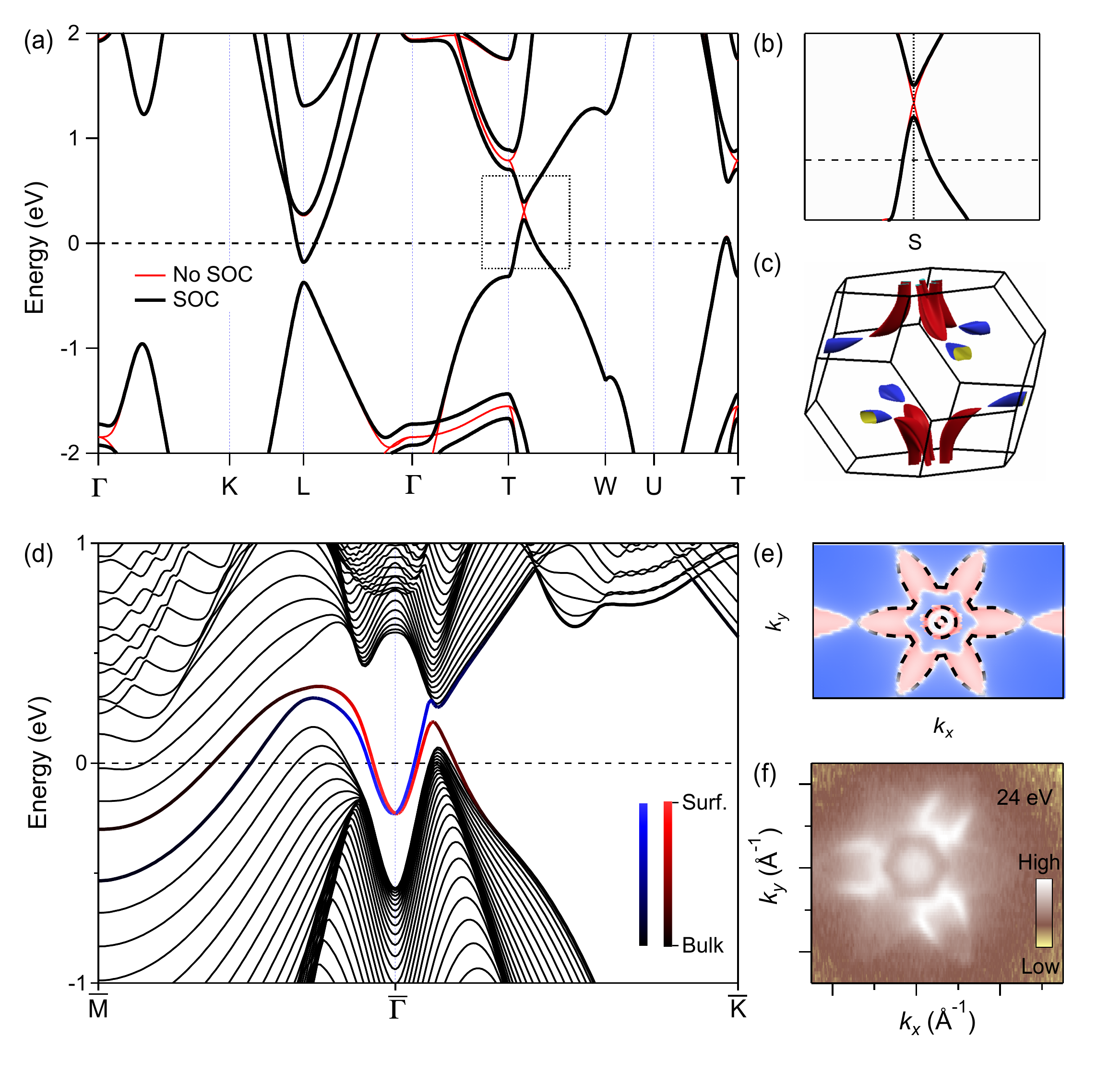}
\end{center}
 \caption{\label{lda} (a) Calculated bulk band structure along high-symmetry lines with SOC (black) and without SOC (red). (b) Zoom-in display of the band structure within the dashed box in (a). (c) Calculated 3D FSs with SOC. (d) Surface and bulk bands along $\bar M-\bar\Gamma-\bar K$ from slab calculations with SOC. The surface components are shown with blue and red colors, while the bulk components are shown with black color. (e) FSs projected onto the (111) surface BZ from tight-binding calculations. The dashed black lines indicate the surface state FSs. (f) FSs recorded with photon energy of 24 eV showing a six-fold flower-like surface state FS. Three branches with high spectral intensity are a combination of the bulk states and surface states, while the other three branches with weak intensity are mainly from the surface states. The two concentric surface state FSs at the BZ center are not distinguishable in the synchrotron ARPES experiments. }
\end{figure}

We perform first-principles calculations to confirm the topological nature of the band structure. The calculations show that there is a band inversion between 4\textit{p} and 5\textit{s} states at the $T$ point induced by crystal field. As seen in Fig.~\ref{lda}(a), the band inversion leads to a band crossing slightly above $E_\mathrm{F}$ along $T-W$. We denote the band crossing point along $T-W$ as the $S$ point [Fig.~\ref{lda}(b)]. In the absence of spin-orbit coupling (SOC), the band crossing forms a Dirac nodal line in the three-dimensional (3D) BZ, as shown in Fig.~S4 in the Supplemental Materials. When the SOC is turned on, the nodal line is fully gapped, leading to a direct energy gap between the valence and conduction bands at every momentum in the whole BZ. In analogy to 3D band insulator, we can derive the $Z_2$ topological invariant $\nu_0$, which distinguishes the strong topological insulator (TI) ($\nu_0$ = 1) from weak TI and ordinary insulator ($\nu_0$ = 0). Since the lattice has inversion symmetry, $\nu_0$ can be calculated from the parity eigenvalues of all valence bands at eight time-reversal invariant momenta (TRIMs), i.e., $\Gamma$, $T$, three $L$ and three $X$ points. The calculation of $Z_2$ invariant of grey arsenic is shown in Table I. The $\nu_0 = 1$ indicates that the grey arsenic belongs to a strong TI~\cite{KanePRB2007}, which is consistent with the previous calculations~\cite{BenedekPRB2012}. In Fig.~\ref{lda}(d), we present the slab calculations of surface and bulk band structures, which reproduce well the experimental results. Around $\bar\Gamma$, the surface bands follow the NFE behavior well. As the surface bands approach the $S$ point, the two branches connect to the bulk conduction and valence states, respectively. The lower branch of the surface bands bends downwards as it merges into the bulk valence states. This leads to a flower-like FS formed by the surface bands, which encloses the projected FS continuum of the bulk valence states, as displayed in Fig.~\ref{lda}(e). This is confirmed by the FS mapping in Fig.~\ref{lda}(f), which shows a six-fold surface state FS enclosing a three-fold bulk state FS. In total, there are three surface state FSs enclosing the TRIM $\bar\Gamma$ point, which supports a nonzero Berry's phase. This is distinguished from the trivial Rashba-split states, in which two concentric spin-polarized FSs enclose $\bar\Gamma$. 

\begin{figure}[!htp]
\begin{center}
\includegraphics[width=0.5\textwidth]{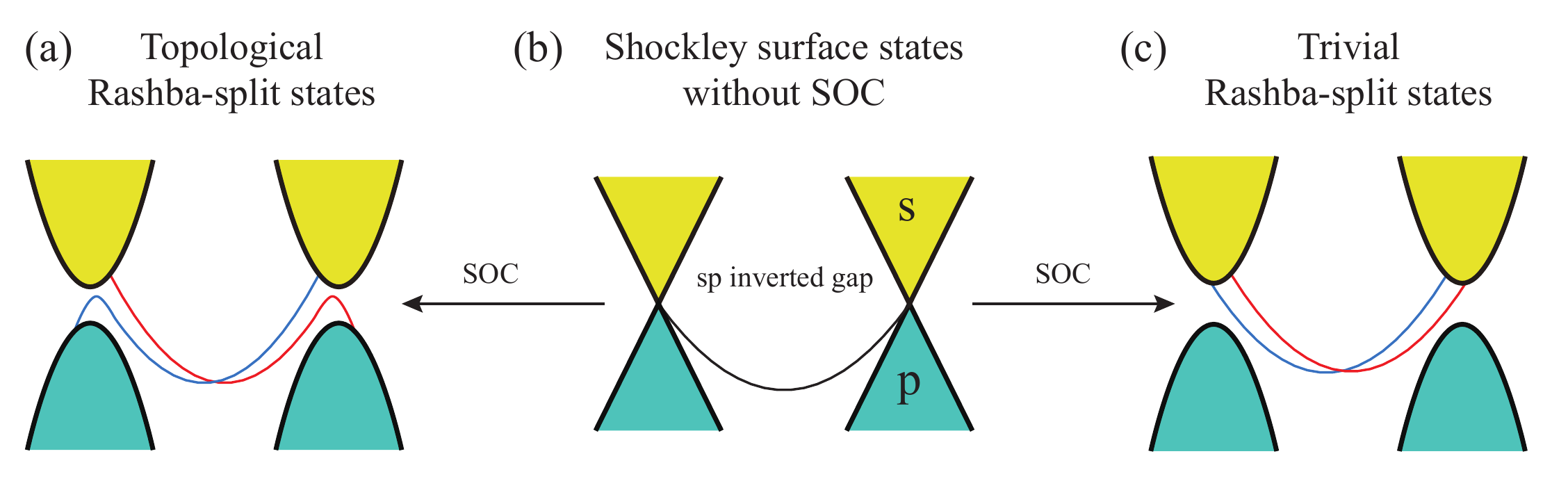}
\end{center}
 \caption{\label{shockley} Evolution of the Shockley states with SOC. (b) Shockley surface states emerging within an \textit{s}-\textit{p} inverted gap of the bulk states caused by crystal field. The surface band with a NFE dispersion connects to the bulk band crossing points without SOC. (a) and (c) Topologically nontrivial and trivial Rashba-split surface states, respectively, which are evolved from the Shockley states in (b) after introducing SOC, depending on the topology of the band structures. }
\end{figure}

The band topology of grey arsenic is essentially the same as the well-known strong TIs Bi$_2$Se$_3$ family. The surface states on As(111) can be regarded as a heavily deformed Dirac cone. However, the origin of the surface states on As(111) is distinguished from the Dirac cone in the Bi$_2$Se$_3$ family. The band inversion in the Bi$_2$Se$_3$ family is induced by the SOC, and thus the surface Dirac cone appears only when the SOC is included \cite{ZhangNP2009}. In contrast, the band inversion in grey arsenic is caused by crystal field and the surface states persist even without SOC \cite{BenedekPRB2012}, which means the surface states on As(111) are of Shockley type. In the absence of SOC, the Shockley states connect to the bulk states at the band crossing points, as shown in Fig.~\ref{shockley}(b). When the SOC is included, the bulk states are gapped at the band crossing points. The SOC has two effects on the Shockley states: causing Rashba splitting and inducing the evolution to topologically trivial or nontrivial surface states, depending on the topology of the bulk band structure [Figs.~\ref{shockley}(a) and \ref{shockley}(c)]. Therefore, the topological surface states on As(111) dramatically deviate from the Dirac-cone-like band dispersion but exhibit Rashba-type band splitting near the TRIM point $\bar\Gamma$. 

The surface Rashba effect originates from the SOC and effective electric field that describes the inversion-symmetry breaking on surface~\cite{Rashba1984,HedegardSS2000,OguchiJPCM2009,KrasovskiiPRB2014}, in which the band splitting itself has no connection with the band topology. Thus it can occur on topologically trivial surface states~\cite{HedegardSS2000}. While in the As case, the spin splitting near $\bar \Gamma$ is dominated by the Rashba effect, but the surface state is topologically entangled and thus its existence is guaranteed by the bulk band topology. In this sense we call this type of surface states on As(111) topological Rashba-split Shockley states. There are emerging proposals for reexamining the topological nature of Shockley states on elemental single crystals. For example, it was recently predicted that in beryllium a band inversion occurs at $\Gamma$ due to the crystal field effect and the derived surface states are topologically nontrivial~\cite{ChenArxiv2016}. As mentioned above, there is also a proposal to explain the Rashba-split Shockley states on noble metals as topologically nontrivial~\cite{FelserNC2015}. However, experimental confirmations of these proposals are difficult, because of either the small energy scale of SOC in beryllium or the surface-to-bulk band connection far above $E_\mathrm{F}$ in noble metals. Instead, we successfully probed the band topology in grey arsenic and our study provides a model for understanding the topological nature of the Shockley states observed in numerous materials.

We acknowledge T. Kondo, K. Kuroda, Z. J. Wang, and X. X. Wu for useful discussions and N. Xu, C. E. Matt, N. C. Plumb, and M. Shi for the assistance in synchrotron ARPES experiments. This work was supported by the Photon and Quantum Basic Research Coordinated Development Program from MEXT, JSPS (KAKENHI Grants No. 25220707, No. 26800165, and No. 15K17675), the National Natural Science Foundation of China (Grants No. 11674371, No. 11474340, No. 11234014, No. 11274359, No. 11422428 and No. 51671193), the Ministry of Science and Technology of China (Grants No. 2013CB921700, No. 2015CB921300, No. 2016YFA0300600, and No. 2016YFA0401000), and the Chinese Academy of Sciences (Grant No. XDB07000000). Part of the calculations were performed on TianHe-1(A), the National Supercomputer Center in Tianjin, China.

\end{document}